\documentclass[aps,reprint,prl]{revtex4-2}
\usepackage{amsmath,graphicx,amssymb}
\usepackage[colorlinks=true,linkcolor=blue,citecolor=red]{hyperref}
\usepackage{siunitx}

\usepackage{amsmath,graphicx,amssymb}
\usepackage[colorlinks=true,linkcolor=blue,citecolor=red]{hyperref}
\usepackage{siunitx}
\usepackage{algpseudocodex}

\DeclareSIUnit{\kohm}{\kilo\ohm}
\DeclareSIUnit{\Mohm}{\mega\ohm}

\DeclareSIUnit{\kohm}{\kilo\ohm}
\DeclareSIUnit{\Mohm}{\mega\ohm}

\begin{document}

\title{Fast and Continuous Detection of Single Microwave Photons via Photo-assisted Quasiparticle Tunneling to a Superconducting Island}

\author{J. Basset}
\email{julien.basset@universite-paris-saclay.fr}
\affiliation{Universit\'e Paris-Saclay, CNRS, Laboratoire  de  Physique  des  Solides,  91405 Orsay,  France}

\author{O. Stanisavljevi\'c}
\affiliation{Universit\'e Paris-Saclay, CNRS, Laboratoire  de  Physique  des  Solides,  91405 Orsay,  France}

\author{J. Gabelli} 
\affiliation{Universit\'e Paris-Saclay, CNRS, Laboratoire  de  Physique  des  Solides,  91405 Orsay,  France} 

\author{M. Aprili} 
\affiliation{Universit\'e Paris-Saclay, CNRS, Laboratoire  de  Physique  des  Solides,  91405 Orsay,  France}

\author{J. Estève}
\affiliation{Universit\'e Paris-Saclay, CNRS, Laboratoire  de  Physique  des  Solides,  91405 Orsay,  France}

\date{\today}

\begin{abstract}
We demonstrate a single-photon detector operating in the microwave domain, based on photo-assisted quasiparticle tunneling events that poison a superconducting island. The detection relies on continuously monitoring the island’s charge parity using microwave reflectometry. This scheme achieves 10\% detection efficiency with sub \SI{50}{\ns} time resolution and short dead time ($\sim \SI{1}{\us}$), for microwave photons at \SI{10}{\GHz}. The detector features three junctions connected to a superconducting island, which together carry out photoelectric conversion and charge readout. 
The enhanced light--matter coupling, crucial to photon to quasiparticle conversion, is provided by a granular aluminum-based high-impedance microwave resonator.
The time resolved detection of itinerant microwave photon opens up new perspectives in quantum sensing, microwave quantum optics and mesoscopic physics.
\end{abstract}
\maketitle

The ability to detect single microwave photons is central to quantum optics, quantum information processing, and quantum thermodynamics~\cite{Blais2021,Pekola2021,Albertinale2021,Casariego2023}. In contrast to the optical domain, where high-efficiency single-photon detectors are mature and widely deployed~\cite{Hadfield2009}, microwave single-photon detection remains challenging because of the extremely low photon energy. Continuous, time-resolved, and efficient detection would substantially expand the capabilities of circuit quantum electrodynamics and microwave quantum optics platforms~\cite{Blais2021}, while enabling new approaches to quantum sensing~\cite{Albertinale2021, Braggio2025} and mesoscopic physics~\cite{Pekola2021}.

Existing microwave photon detectors exhibit important trade-offs. Superconducting-qubit-based detectors achieve high efficiency and photon-number resolution~\cite{Inomata2016,Opremcak2018,Kono2018,Besse2018,Lescanne2020,Dassonneville2020,May2025}, but typically operate in gated mode and suffer from finite dead times. Threshold detectors~\cite{Chen2011,Oelsner2017,Pankratov2025} and bifurcation amplifiers~\cite{Petrovnin2024} provide time-resolved responses but require resetting after switching events, while calorimetric detectors are limited by thermal relaxation times~\cite{Gasparinetti2015,Kokkoniemi2020}. Finally, approaches based on inelastic Cooper-pair tunneling have not yet demonstrated single-photon sensitivity~\cite{Albert2025}, while hybrid semiconductor implementations currently achieve improved efficiency only at the cost of increased dark count rates~\cite{Haldar2024,Oppliger2025}. Consequently, no existing platform simultaneously offers continuous operation with good timing resolution, short dead time and high efficiency at the single-photon level.

In this Letter, we experimentally demonstrate a detector based on photo-assisted quasiparticle tunneling in a superconducting island, enabling continuous, time-resolved detection of single microwave photons with \SI{50}{\ns} timing resolution, \SI{1}{\us} dead time and a detection efficiency of 10\%. In the ready state, the island has even charge parity. Absorption of a photon induces photo-assisted tunneling of a single quasiparticle, switching the parity from even to odd—a mechanism previously employed for infrared photon detection~\cite{Hergenrother1995,Echternach2018}. The resulting parity change is continuously monitored via microwave reflectometry, following the principles of radio-frequency single-electron or Cooper pair transistors~\cite{Tuominen1992,Joyez1994,Amar1994,Schoelkopf1998,Aassime2001,Ferguson2006,Naaman2006}.

Upon photon absorption, our detector produces a real-time electrical pulse analogous to the “click’’ of an optical single-photon detector. Extending photon time tagging to the microwave domain enables measurements of photon correlations~\cite{Bozyigit2011}, linear-optics–based quantum-information protocols~\cite{Casariego2023}, and remote entanglement schemes~\cite{Narla2016}. More broadly, time-resolved microwave photon detection provides a new probe of quantum noise and nonequilibrium dynamics in mesoscopic systems and quantum materials~\cite{Moca2011,Chen2023}.

\begin{figure}[htbp]
	\begin{center}
		\includegraphics[width=8.57 cm]{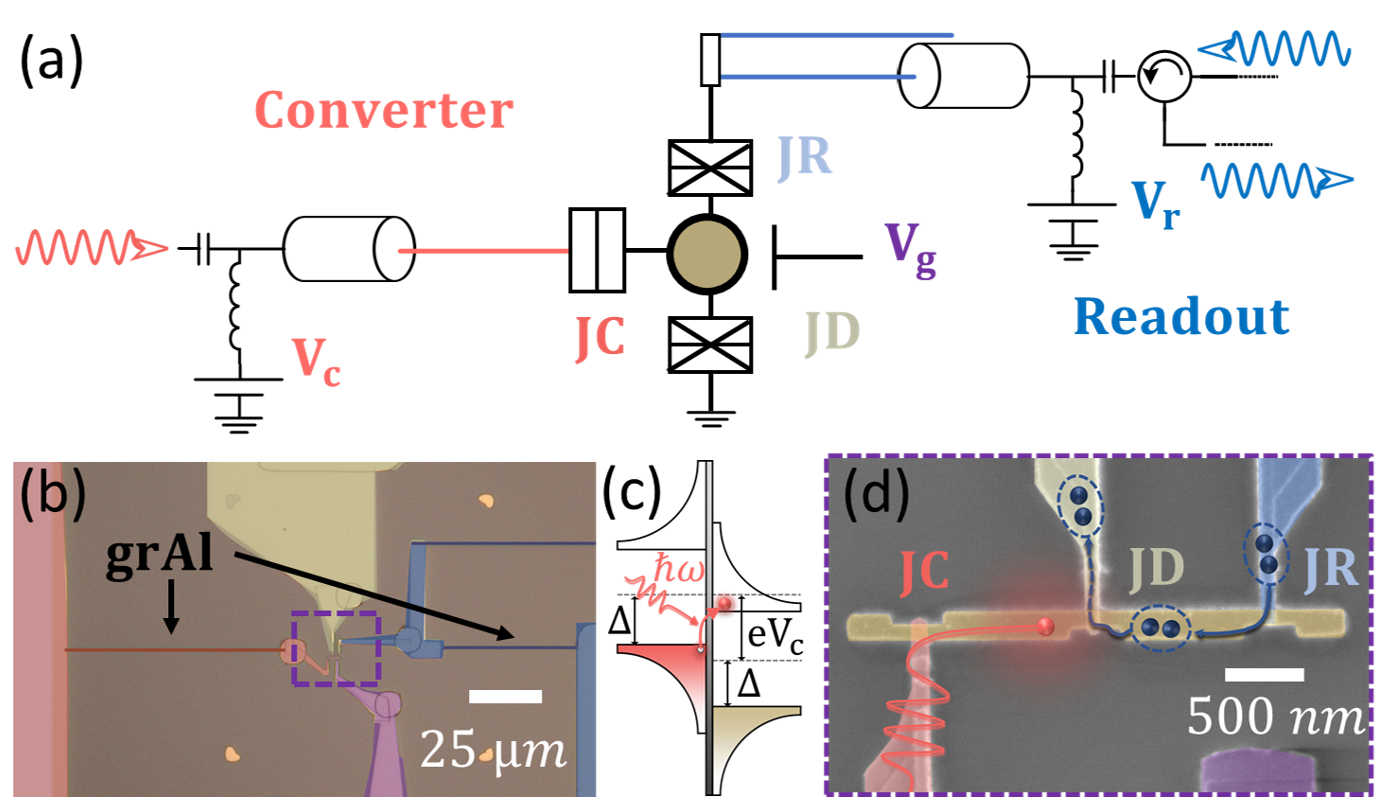}
	\end{center}
	\vspace{-0.4cm}
	\caption{\textbf{Detection principle and device.} 
		(a) Simplified circuit diagram of the detector. An incoming microwave photon at the detector input is absorbed via photo-assisted tunneling into a superconducting island, inducing a quasiparticle-poisoning event. The resulting change in charge parity is continuously monitored through microwave reflectometry at the readout port. Three dc bias voltages ($V_r$, $V_c$, $V_g$) tune the working point of the detector. JR and JD operates with the supercurrent and thus have an extra cross in their representation as compared to JC.  
		(b) Optical micrograph of the device. Both converter and readout resonators are made of \SI{700}{\nm} wide, \SI{20}{\nm} thick  granular aluminum wires with large kinetic inductance ($\SI{0.5}{\nano\henry}/\Box$). The converter resonator is \SI{73}{\um} long and the readout resonator is made of two sections, which are \SI{48}{\um} and \SI{172}{\um} long.
		(c) Energy diagram representation of the microwave photo-assisted tunneling process. 
		(d) Scanning electron micrograph of the superconducting island and the three junctions JC, JR and JD. The gate appears in purple.
	}	
	\label{Figure1}
	\vspace{-0.6cm}
\end{figure}

\textbf{\textit{Principle of the experiment---}}
The operating principle of the detector is illustrated in Fig.~\ref{Figure1}. Details about the device fabrication can be found in the Supplementary Materials~\cite{SM}. A superconducting island is connected through three junctions to the detector input port, the readout port, and ground (Fig.~\ref{Figure1}a). During normal operation of the detector, the readout bias $V_r$ is set to zero, and the gate voltage $V_g$ is adjusted to maximize the critical current through the two junctions (JR, JD) forming the single Cooper pair transistor (SCPT), thereby pinning the island potential to ground. The third junction (JC), referred to as the converter, is biased at $V_c$ just below the onset of quasiparticle tunneling near the superconducting gap. 
In the ready state, the island carries even charge parity. An incoming photon at the detector input can be absorbed through a photo-assisted tunneling (PAT) event at JC, generating a quasiparticle on the island (Fig.~\ref{Figure1}c)~\cite{Devoret1990,Ingold1992}. The resulting charge parity flip modifies the impedance of the readout circuit until the quasiparticle leaves the island, after which the detector self-resets on a timescale of $\sim \SI{1}{\us}$. Monitoring the reflection of a microwave tone from the readout circuit yields a square-pulse signal marking each photon-detection event.

Central to the detector’s performance is the use of granular aluminum (grAl) to realize high-impedance microwave resonators~\cite{Maleeva2018}. Our device incorporates two grAl quarter-wavelength resonators, used respectively for efficient photon-to-electron conversion and for charge readout (see Supplementary Materials~\cite{SM} for details on the microwave design).
The first quarter-wavelength grAl resonator precedes JC and provides impedance matching at $\nu \approx \SI{10}{\GHz}$ from the \SI{50}{\ohm} detector input to the high impedance junction. Unity photon-to-quasiparticle conversion efficiency is achieved when the resonator coupling rate $\kappa_c$ matches the photon absorption rate $\kappa_j$ due to photo-assisted tunneling (PAT) through JC. We previously demonstrated an efficiency exceeding 80\%~\cite{Stanisavljevic2024} in a single junction device, and a comparable result was recently obtained with a double quantum dot~\cite{Oppliger2025}. 

For a junction coupled to a single mode resonator, $\kappa_j$ scales with the junction conductance and the Franck-Condon factor $\lambda^2  \exp(-\lambda^2)$, where the dimensionless light-matter coupling constant $\lambda = \sqrt{\pi Z_c/R_K}$ depends on the ratio between the mode’s characteristic impedance $Z_c$ and the quantum of resistance $R_K = h/e^2$. The Franck-Condon factor for elastic tunneling is simply $\exp(-\lambda^2)$. Large $\lambda$ therefore enhances the PAT-to-elastic rate ratio and suppresses associated dark counts; our grAl resonator achieves $\lambda=0.63$. The converter junction, realized via triple oxidation, has a normal-state resistance of \SI{6.4}{\Mohm}. The other modes in the circuit, including the higher order modes of the converter resonator and the modes of the Cooper pair transistor, also induce a dynamical Coulomb blockade and reduce the PAT rate. Including all these effects, as detailed in the Supplementary Materials~\cite{SM}, we evaluate a rate of $\kappa_j \approx 2\pi\times\SI{10}{\MHz}$, below the coupling rate, which we estimate to $\kappa_c \approx 2\pi\times\SI{70}{\MHz}$. This rate mismatch limits the quantum efficiency around $4\kappa_c\kappa_j/(\kappa_c+\kappa_j)^2\approx 50\%$, suggesting that further engineering could improve performance.

The quasiparticle generated by the PAT process flips the island's charge parity from even to odd. This parity switching suppresses the critical current of the SCPT, changing its microwave impedance, thus shifting the resonant frequency of the readout resonator. The resonator, also made of grAl, is designed to have a smaller light-matter coupling strength around $0.2$ and a resonant frequency near \SI{5.2}{\GHz}. The two SCPT junctions, JR and JD, have normal-state resistances summing up to \SI{136}{\kohm}. The island, fabricated from a thin (\SI{15}{\nm}) aluminum layer, has an enhanced gap to suppress quasiparticle poisoning through JR and JD, which connect to thicker Al leads~\cite{Aumentado2004}.

The reflection coefficient $S_{11}$ upon the readout port is monitored with a probe tone nearly resonant with the readout resonator. Two types of measurements were conducted: one employing a Josephson Parametric Amplifier (JPA) on the returning path of the readout signal, and one without (see Supplementary Materials~\cite{SM} for the detailed microwave setup). The probe amplitude is adjusted to obtain two well-separated signals in the IQ plane corresponding to each charge parity in less than \SI{100}{\ns} integration time. This behavior contrasts with the standard operation of a radio-frequency single-electron transistor (RF-SET) operation, which detects small charge offsets producing only minor $S_{11}$ variations~\cite{Schoelkopf1998,Aassime2001}. Without JPA, we achieve a sensitivity of $6\times10^{-6}\,e/\sqrt{\rm Hz}$, comparable to state-of-the-art RF-SETs~\cite{Brenning2006}. 

Instead, the readout is similar to previous real time parity detection in SCPTs ~\cite{Ferguson2006,Naaman2006} and in charge sensitive transmon qubits ~\cite{Serniak2019,Amin2024}. In our case, the charge detector operates in the Cooper pair box regime, chosen to maximize readout speed. The detector self-resets to the even-parity state within \SI{1}{\us} as previously observed~\cite{Ferguson2006}. Charge-parity readout can be performed continuously or in pulsed mode, in which case a readout pulse follows the signal pulse at the detector input. Optimization of the detector performance primarily involves tuning the readout power and integration time to balance detection efficiency and dark counts. Depending on the intended application, one can either reduce detection errors by increasing the readout power—at the expense of higher dark counts—or minimize dark counts at the cost of reduced detection efficiency (see Supplementary Materials~\cite{SM}).

\begin{figure}[htbp]
	\begin{center}
		\includegraphics[width=8.57 cm]{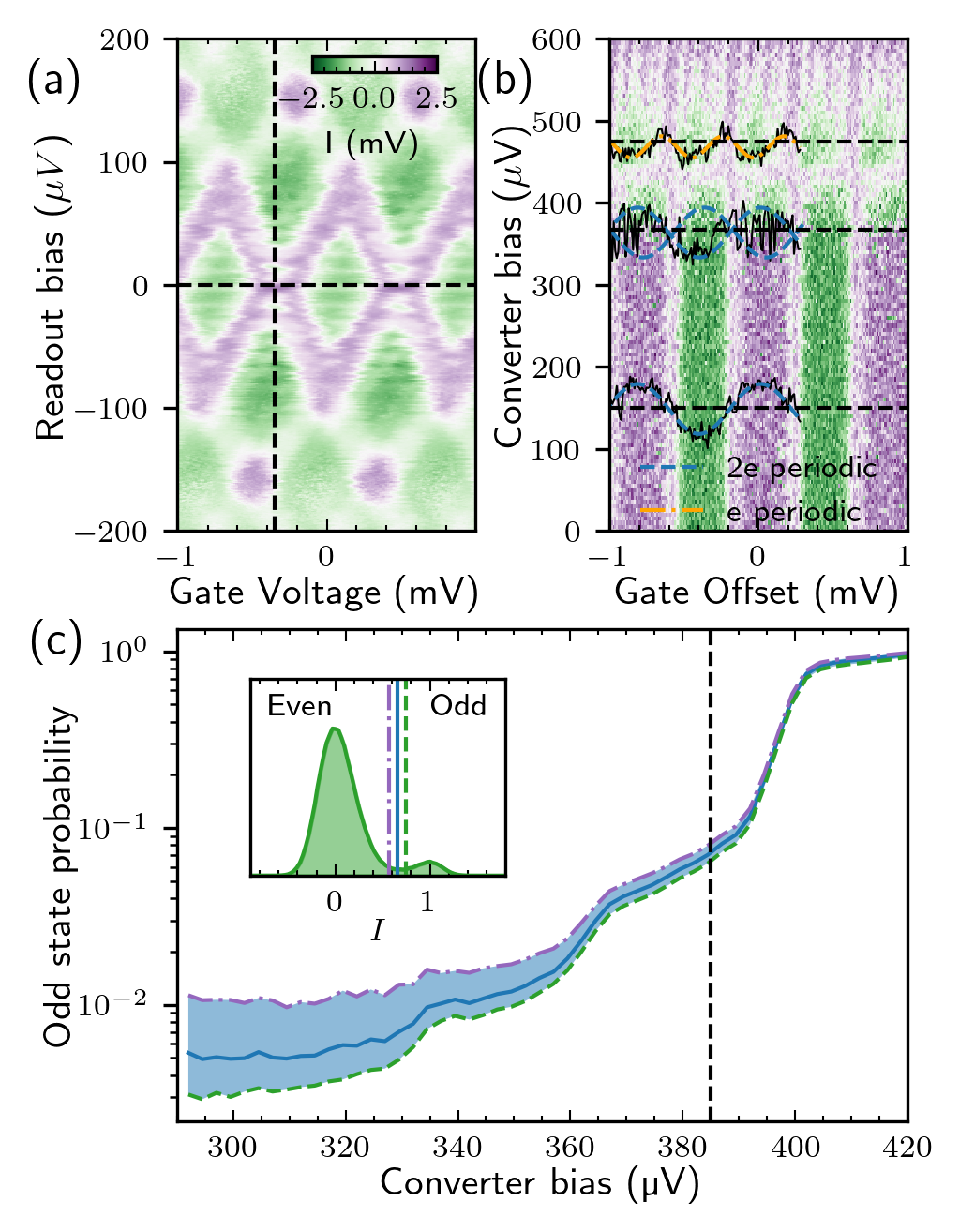}
	\end{center}
	\vspace{-0.4cm}
	\caption{\textbf{Charge stability with voltages.} 
		(a) Dependence of one quadrature, $I$, of the reflected readout signal on gate voltage and readout bias at zero converter bias. The observed Coulomb diamonds are $2e$ periodic and delimit stable charge states with a well defined number of Cooper pairs. The crosshair indicates the detector operating bias point. 
		(b) Dependence of the readout signal at zero readout bias on gate voltage and converter bias. Selected gate voltage line cuts are overlaid with \(2e\) (blue) and \(e\)-periodic (yellow) sinusoidal curves, illustrating the transition from a $2e$ to $e$ periodic response. Color scale is the same as in (a).   
		(c) Probability of the odd charge state as a function of converter bias $V_c$ in absence of photon irradiation. The inset shows the histogram of the readout signal at $V_c=\SI{385}{\uV}$ integrated for \SI{100}{\ns} using the JPA setup. The measured quadrature is rescaled and offset such that the two maxima are located at 0 and 1. The three vertical lines indicate the thresholds used to extract the odd-state probability $p_1$ shown in the main plot. Each curve corresponds to the threshold lines indicated in the inset.
	}	
	\label{Figure2}
	\vspace{-0.4cm}
\end{figure}

\textbf{\textit{Experimental Results---}}
Figure~\ref{Figure2}a shows one quadrature of the reflected readout signal as a function of gate and dc bias across the SCPT. Coulomb diamonds are observed, periodic in gate voltage with a period corresponding to a $2e$ change in island charge. In the presence of an additional quasiparticle of charge $e$, the entire pattern shifts by half a period. From the diamond height, the charging energy of the SCPT is estimated to $E_C\approx \SI{20}{\micro\eV}$.
The charge parity readout is operated at $V_r=0$, while $V_g$ is adjusted to maximize the SCPT critical current (see Fig.~\ref{Figure2}a). To maintain operation at this optimal point during the detector operation, the gate voltage is periodically adjusted every few seconds to compensate for slow charge-offset drift and for abrupt charge jumps of $e$ occurring on minute timescales. The gate dependence of the reflection coefficient $S_{11}$ is recorded and compared to a reference waveform to extract the offset. When a voltage is applied to JC, an additional correction is applied based on the known capacitive lever arm, calibrated in a separate measurement.

We first examine the rate at which quasiparticles enter the island in the absence of any incident signal on the detector. Figure~\ref{Figure2}b shows the evolution of one quadrature of the reflected signal $S_{11}$ as a function of the gate offset from the optimal point and the voltage $V_c$ across the converter junction JC. As $V_c$ approaches the superconducting gap, a transition from $2e$- to $e$-periodic gate modulation is observed, indicating the onset of quasiparticle poisoning. Each data point corresponds to a \SI{1}{\micro\second}-long measurement, with the gate voltage serving as the fast scanning axis. When the average time between two parity jumps exceeds the measurement time but remains shorter than the scan duration, random $e$-period shifts appear in the gate response (see overlaid traces in Fig.~\ref{Figure2}b). When the poisoning rate surpasses the measurement bandwidth, the island switches rapidly between even and odd parity states, producing an $e$-periodic signal even within a single trace.

With the detector tuned to the optimal gate bias, we record a series of $S_{11}$ measurements as a function of the converter bias $V_c$. Consecutive measurements are separated by \SI{20}{\micro\second} to minimize spurious parity jumps due to the readout tone. We determine the optimal phase rotation that encodes the charge parity onto a single quadrature at optimal $V_g$ and construct the corresponding histograms as exemplified in the inset of Fig.~\ref{Figure2}c. At low $V_c$, the histograms exhibit a single Gaussian peak, while as $V_c$ approaches the superconducting gap, they evolve into two well-separated peaks corresponding to the even and odd charge-parity states. The double-peak structure is not accurately described by a simple sum of two Gaussian functions; therefore, we employ a thresholding method to extract the probabilities $p_0$ and $p_1$ that the charge parity is even or odd, respectively. Because the two distributions partially overlap, the readout error probabilities $\epsilon_i$, corresponding to assigning the opposite outcome $\bar{i}$ when the true parity is $i$, remain significant. We develop a minimal model that includes parity switching during the measurement and estimate $\epsilon_0 +\epsilon_1 \approx 0.1$, as detailed in the Supplementary Materials~\cite{SM}.

The evolution of $p_1$ with $V_c$ extracted from the histograms is shown in Fig.~\ref{Figure2}c. The sharp increase beginning near \SI{400}{\micro\volt} marks the onset of elastic quasiparticle tunneling. We  attribute the step observed around \SI{370}{\micro\volt} to inelastic tunneling processes involving the absorption of stray photons in a parasitic mode at \SI{13}{\giga\hertz}. The probability $p_1$ increases with higher readout power (see Supplementary Materials~\cite{SM}); however, the use of a JPA enables operation at sufficiently low power to suppress this effect. We also observe that $p_1$ varies between cooldowns and depends on the charge environment as it shows clear correlation with the evolution of the periodically measured charge offset.

\begin{figure}[htbp]
	\begin{center}
		\includegraphics[width=8.57cm]{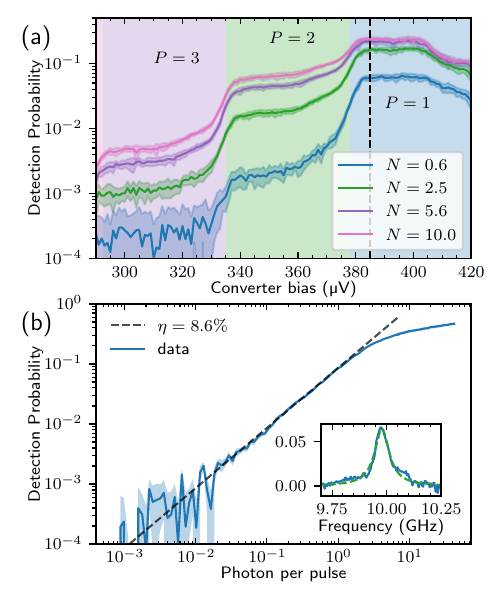}
	\end{center}
	\vspace{-0.4cm}
	\caption{\textbf{Photo-assisted Charge Parity Jump.} 
		(a) Increase in the odd-charge-state probability, $p_1(N) - p_1(0)$, after sending a pulse containing $N$ photons at \SI{10}{\giga\hertz} to the detector input. The colored regions, each of width $h \nu/e$, indicate the voltage ranges where $P$-photon-assisted tunneling processes are allowed by energy conservation. The shaded area represents the $1\sigma$ statistical uncertainty.
		(b) Detection probability $p_1(N)-p_1(0)$ as a function of $N$ at $V_c=\SI{385}{\uV}$ (vertical line in a). The data are obtained with the JPA. Inset: Frequency dependence of the detection probability measured at $N=0.8$; the Lorentzian fit yields a FWHM linewidth of \SI{82}{\mega\hertz}.
	}	
	\label{Figure3}
	\vspace{-0.6cm}
\end{figure}

We next repeat the experiment by sending a \SI{18}{\nano\second} (FWHM) pulse at frequency $\nu$ onto the detector input, containing on average $N$ photons per pulse, just before the readout pulse. The detection probability, defined as $p_1(N) - p_1(0)$, is plotted in Fig.~\ref{Figure3}a as a function of $V_c$. Distinct steps appear, separated by $h\nu / e = \SI{41}{\micro\volt}$, corresponding to PAT processes involving the absorption of $P = 1$, 2, and 3 photons~\cite{Stanisavljevic2024}. We fix $V_c = \SI{385}{\micro\volt}$, corresponding to the first step ($P=1$) below the gap and record $p_1(N)$ as a function of $N$, as shown in Fig.~\ref{Figure3}b. The calibration of $N$ is detailed in the Supplementary Materials~\cite{SM} and relies on an independent estimation of the total attenuation between the microwave source and the detector input.

At low photon numbers ($N\leq 1$), $p_1(N)$ increases linearly with $N$, yielding an operational detection efficiency $\eta=8.6\%$. The efficiency is almost independent of the chosen threshold, which is set at the minimum between the two peaks of the histogram measured at $N=0$. The value of $\eta$ can be expressed as $\eta={\rm QE}(1-2p_{\rm odd})(1-(\epsilon_{0}+\epsilon_{1}))$, where $\rm QE$ is the quantum efficiency (photon to charge conversion efficiency) and $p_{\rm odd}$ is the probability that the island is in an odd-parity state when the photon arrives. Correcting for these effects ($\epsilon_{0}+\epsilon_{1}=0.1$, $p_{\rm odd}=0.145$) gives an estimated ${\rm QE} \approx 13.5\%$. The detector bandwidth is $\kappa = 2\pi \times \SI{82}{\MHz}$ (see inset in Fig.~\ref{Figure3}b), a large number compared to sub-MHz qubit experiments~\cite{Balembois2024} which we can tolerate thanks to our fast charge readout. A possible explanation for the reduced quantum efficiency relative to the expected value is charge noise in the SCPT, which induces phase noise in the converter mode. 
Performing the same measurement without the JPA yields a comparable detection efficiency of 9.1\%, albeit with a higher dark count probability.  

\begin{figure}[htbp]
	\begin{center}
		\includegraphics[width=8.57 cm]{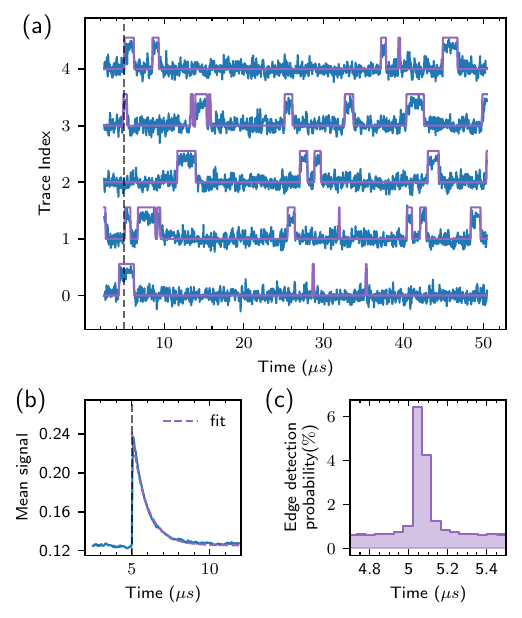}
	\end{center}
	\vspace{-0.8cm}
	\caption{\textbf{Continuous photon detection} (a) Time traces of the readout signal binned with a resolution of \SI{48}{\nano\second}. The reconstructed charge parity identified by the Viterbi algorithm are shown as purple traces. An incoming pulse with 1.12 photon is applied at $t = \SI{5}{\micro\second}$.
		(b) Averaged trace obtained from 4096 repetitions of the measurement in (a), showing the pulse arrival and subsequent recovery of the detector.
		(c) Probability of detecting an even to odd parity jump as a function of time near $t = \SI{5}{\micro\second}$. 
	}	
	\label{Figure4}
	\vspace{-0.6cm}
\end{figure}

A key advantage of our detection scheme is its ability to operate in continuous mode. Figure~\ref{Figure4}a shows time traces of the $S_{11}$ quadrature encoding the charge parity, sampled with a binning time of \SI{48}{\nano\second}. The traces exhibit discrete switching between low and high levels, corresponding to the even and odd charge-parity states, respectively. During the measurement, a single pulse containing on average 1.1 photons is applied to the detector input at $t = \SI{5}{\micro\second}$. The averaged trace shown in Fig.~\ref{Figure4}b displays a sharp transition from even to odd parity coinciding with the pulse arrival, followed by an exponential recovery to the even (ready) state with a characteristic time constant $\tau = \SI{0.94}{\micro\second}$. The relaxation time $\tau$ varies slightly between cooldowns and depends weakly on $V_c$ and the readout power, but consistently remains around \SI{1}{\micro\second} or below.

To reconstruct the time evolution of the charge parity from the measured time traces, we employ a maximum-likelihood reconstruction based on the Viterbi algorithm (see Supplementary Materials~\cite{SM}). The algorithm requires the relaxation and poisoning rates, $\gamma_{01}$ and $\gamma_{10}$, as well as the readout histograms. The rates are determined from the steady-state average parity $\gamma_{10}/(\gamma_{10} + \gamma_{01})$ and the measured relaxation time $\tau^{-1} = \gamma_{10} + \gamma_{01}$. The detection histograms are modeled as two Gaussian distributions. The reconstructed trajectories are overlaid with the measured time traces in Fig.~\ref{Figure4}a. 
A rising front (even-to-odd parity transition) identified by the algorithm marks a detection event with a time resolution set by the $\SI{48}{\ns}$ binning time. The single-photon detection efficiency is $\eta=10\%$ and the measured dark count rate of the detector is $\gamma_{10} = \SI{137e3}{\per\second}$.
From the analysis of numerically generated time traces, we estimate the probability to miss an event to 10\% and the probability of false detection to 1.5\%. 

Figure~\ref{Figure4}c shows the probability that an event is detected close to the pulse arrival time. No measurable timing jitter is observed beyond the binning resolution. This is consistent with the expected temporal limit set by the converter cavity linewidth, which determines the rise time of the photo-assisted tunneling (PAT) probability. Based on our estimate of $\kappa_c$, we expect only marginal temporal broadening of the pulse width from \SI{18}{\ns} to \SI{20}{\ns}. The readout response can, in principle, follow parity changes almost instantaneously; in practice, its speed is constrained by the detuning induced by a parity jump in the readout resonator, which can be significant.

Finally, we have measured the evolution of the poisoning rate $\gamma_{10}$ at $V_c=\SI{385}{\uV}$ as a function of temperature. The data shown in the Supplementary Materials~\cite{SM} are compatible with a quantum efficiency of 56\%. This suggests that the quantum efficiency measured at a fixed frequency is likely limited by random fluctuations of the detection band due to charge noise. Under these assumptions, the observed dark count rate corresponds to a residual mode population of $3\times 10^{-3}$. State-of-the-art filtering can suppress populations below $10^{-5}$ at \SI{10}{\GHz}~\cite{Balembois2024}. Whether such suppression can be achieved in photo-assisted–junction detectors remains an open question.

As such, our detector already distinguishes itself through its continuous and time-resolved mode of operation compared with state-of-the-art itinerant microwave photon detectors~\cite{Besse2018,Lescanne2020,Kokkoniemi2020,Albertinale2021,Haldar2024}. In comparison with photo-assisted-tunneling–based semiconductor detectors, our approach achieves a substantial detection efficiency in the photon-counting regime~\cite{Haldar2024,Oppliger2025}. Relatively modest device-level optimizations, such as improving the coupling rate and/or tuning the junction transparency, could further enhance the efficiency~\cite{Stanisavljevic2024}.

Achieving near-unity detection efficiency together with thermally limited dark counts will enable real-time photon-correlation measurements, opening new routes for quantum information processing with microwave photons analogous to those developed in optics. 
We estimate that a detector with 80\% efficiency and a 100 kHz dark-count rate would allow a seond-order correlation function $g(2)(\tau)$ experiment with 100 ns resolution and 100 kHz repetition rate to reach a signal-to-noise ratio of order unity in $\approx80$~ms. In contrast, a linear detection chain with input-referred noise of two photons would require $\approx160$~s, corresponding to a $2000$-fold reduction in measurement time.

More broadly, quantum-noise measurements—so far primarily performed using linear amplifiers—could greatly benefit from time-tagged detection, providing direct access to electron correlations and nonequilibrium dynamics in mesoscopic nanostructures and quantum materials~\cite{Clerk2010,Moca2011,Chen2023}.

\textbf{\textit{Acknowledgements---}}
We thank P. Bertet, E. Flurin, A. May, C. Di Giorgio for stimulating discussions, B. Reulet for lending us the JPA and R. Weil for technical support during microfabrication. We acknowledge financial support from the ANR (ANR-21-CE47-0010 KIMIDET project), from the France 2030 plan under the ANR-22-PETQ-0003 RobustSuperQ grant, from the Région Ile-de-France in the framework of DIM QuanTiP (Quantum Technologies in Paris Region). This work was carried out as part of the NanoLPS micro-nanotechnology platform of the french RENATECH+ network.

\vspace{-10mm}
\bibliography{biblio}

\clearpage
\onecolumngrid
\appendix

\appendix

\renewcommand{\thefigure}{S\arabic{figure}}
\renewcommand{\thetable}{S\arabic{table}}
\renewcommand{\theequation}{S.\arabic{equation}}
\setcounter{figure}{0}
\setcounter{table}{0}
\setcounter{equation}{0}

\begin{center}
	\huge{Supplemental material}
\end{center}

\section{Sample Fabrication}
The sample is fabricated using a combination of optical and electron-beam lithography. In the first step, electron-beam lithography defines \SI{700}{nm}–wide, \SI{20}{nm}–thick granular aluminium (grAl) nanowires, deposited by e-gun evaporation with a small partial pressure of oxygen and patterned by lift-off. A second electron-beam lithography step is then used to define the three tunnel junctions and the superconducting island. Aluminium is first evaporated at normal incidence to a thickness of \SI{15}{nm}, followed by a controlled oxidation for 15 minutes at \SI{10}{mbar}. A second evaporation at a positive angle deposits \SI{30}{\nm} of Al to form the more transparent junctions (readout and drain), followed by two additional oxidation steps separated by a \SI{0.2}{\nm} Al flash, and finally a third evaporation of \SI{65}{nm} Al at a negative angle. All evaporations are performed without breaking vacuum and followed by lift-off. In a final step, optical lithography defines the aluminium transmission lines and contact pads connecting the grAl resonators and the junction region, deposited by a \SI{120}{\nm} Al evaporation.

\section{Microwave Design}
\label{MicrowaveDesign}
\begin{figure}[htbp]
	\begin{center}
		\includegraphics[width=15 cm]{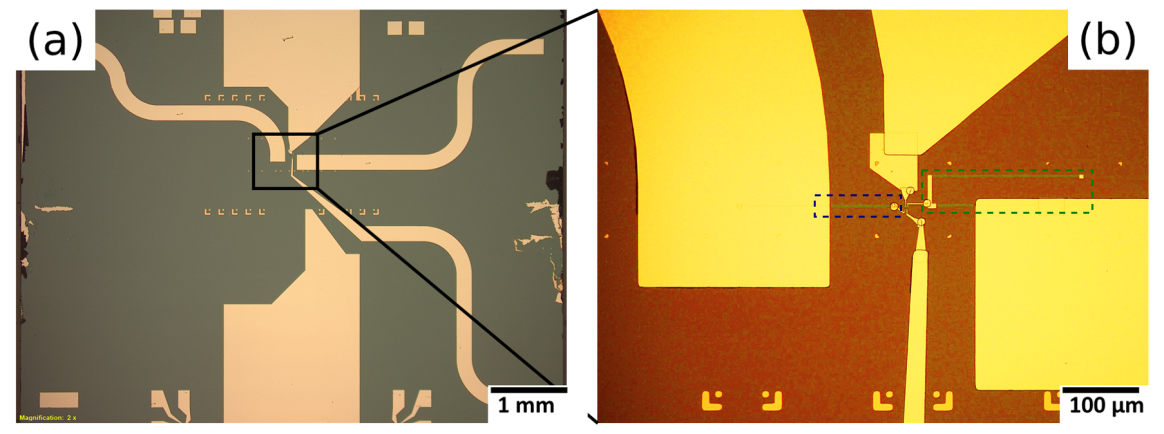}
	\end{center}
	\vspace{-0.4cm}
	\caption{(a–b) Optical microscope images of the sample at increasing magnifications. In (b), the contrast and brightness have been adjusted to enhance the visibility of the grAl wires. The blue (green) dashed rectangle indicates the converter (readout) resonator.
	}	
	\label{SI_Fig_Sample}
\end{figure}

The sample consists of two quarter-wavelength resonators, used respectively for photon-to-electron conversion and charge readout. In the main text, these are referred to as the converter and readout resonators. Both are fabricated from a narrow (\SI{700}{\nm}) grAl wire directly connected to a much wider (\SI{220}{\um}) aluminum microstrip waveguide with \SI{50}{\ohm} characteristic impedance, as shown in Fig.~\ref{SI_Fig_Sample}. This direct galvanic connection enables dc biasing of the junctions using either a bias-tee or a frequency diplexer located outside the sample.

The grAl film has a sheet resistance of \SI{800}{\ohm}, corresponding to a sheet inductance of \SI{0.53}{\nano\henry} and a linear inductance of \SI{0.76}{\milli\henry\per\meter} for the wire. Electromagnetic simulations performed with Sonnet yield a linear capacitance of \SI{42}{\pico\farad\per\meter}. From these values, we extract a characteristic impedance of the grAl microstrip waveguide of $Z_c=\SI{4.3}{\kilo\ohm}$ and a phase velocity of \SI{5.6d6}{\meter\per\second}. Owing to the large impedance mismatch with the \SI{50}{\ohm} line, the reflection coefficient at this interface is close to $-1$.

On the opposite end, the converter resonator is terminated by a tunnel junction biased close to the superconducting gap, which behaves predominantly as a capacitor of a few femtofarads. The associated impedance is comparable to the characteristic impedance of the grAl line and therefore cannot be neglected. The junction is placed at the end of the resonator to maximize the coupling between the resonant mode and tunneling electrons.

We model the device as a one-dimensional transmission line terminated on one side by a \SI{50}{\ohm} load and on the other side by a capacitance. This simplified model enables analytical calculations of the mode structure. A resonator length of \SI{73}{\um} was chosen to place the fundamental mode near \SI{10}{\giga\hertz}, including the effect of a junction capacitance of a few femtofarads. The precise capacitance value (\SI{4}{\femto\farad}) is adjusted a posteriori to match the measured resonance frequency. The model then yields the resonance frequencies, characteristic impedances, and quality factors for the converter modes, as summarized below:
\begin{center}
	\begin{tabular}{c c c}
		Frequency (GHz) & Impedance (\si{\kilo\ohm}) & Q \\
		9.52 & 3.28 & 55 \\
		41.4 & 116 & 153 \\
		78.5 & 18 & 278 \\
	\end{tabular}
\end{center}
The quality factors cannot be reliably predicted by the model because it accounts only for the impedance step and neglects the geometrical size mismatch, which also enhances the reflection coefficient. Experimentally, we measure a linewidth of $\kappa_c = 2\pi \times \SI{70}{\MHz}$ (corresponding to $Q=140$), which is significantly smaller than the simulated value. We therefore use the measured linewidth when estimating the photon-to-electron conversion efficiency.

The readout resonator is divided into two sections connected by an aluminum pad. The two segments have lengths of 48 and \SI{172}{\um}, respectively. By coupling the readout junction to an intermediate point along the resonator rather than at its end, the coupling between the resonant mode and the junction is reduced. In this configuration, the voltage at the junction is approximately half of that at the resonator termination. The simple transmission-line model predicts a fundamental quarter-wavelength mode at \SI{6.4}{\giga\hertz}. The experimentally observed resonance at \SI{5.5}{\giga\hertz} is reproduced by introducing an additional capacitance of \SI{10}{\femto\farad} to account for the junctions and the Al patch between the two nanowire sections. A more comprehensive model that incorporates both the Josephson energy and the charging energy of the island is presented in section \ref{CPB} below.

\section{Charge Readout of the Island}
\label{CPB}

\begin{figure}[htbp]
	\begin{center}
		\includegraphics{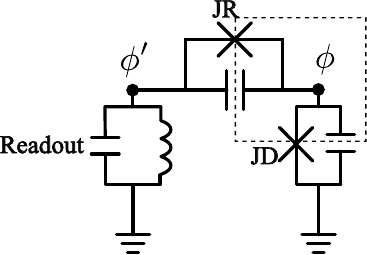}
	\end{center}
	\vspace{-0.4cm}
	\caption{Lumped element model of the charge readout circuit. The island is indicated by the dashed rectangle. The readout resonator is modeled by the parallel $LC$ circuit on the left. The converter junction and the converter resonator are not drawn.}	
	\label{SI_Fig_Readout}
\end{figure}
The charge readout circuit can be modeled as shown in Fig.~\ref{SI_Fig_Readout}. The two dynamical variables, $\phi$ and $\phi'$, denote respectively the superconducting phase of the island and the readout resonator mode. The Hamiltonian of the circuit reads
\begin{equation}
	H = \hbar\omega a^\dagger a + 4E_C(n-n_0)^2 -E_{JD} \cos \phi + 4i \epsilon E_C \lambda^{-1} \sin kl \, (n-n_0)(a-a^\dagger)  - E_{JR} \cos \big[\phi - 2\lambda \sin kl \,(a+a^\dagger)\big] \label{eq.readout_H} 
\end{equation}
The first three terms describe the uncoupled resonator mode at $\omega\approx 2\pi \times \SI{5.5}{\GHz}$ and island Hamiltonians, while the remaining terms account for their coupling mediated by junction JR. The offset charge $n_0$ is controlled by the gate voltage. The island charging energy is $E_C/h=\SI{4.6}{\GHz}$, extracted from Coulomb-diamond measurements near the superconducting gap.
The tunnel resistances of the two junctions JR and JD in series is \SI{136}{\kilo\ohm}, yielding Josephson couplings $E_{JD}/h=E_{JR}/h=\SI{2.3}{\GHz}$. Neglecting any renormalization due to capacitive effects, the dimensionless factor $\lambda=2(Z_c/R_K)^{1/2}=0.8$ sets the amplitude of the flux $\phi'$, expressed in units of the superconducting flux quantum, in terms of the creation and annihilation operators of the readout mode. The factor $\sin kl$ accounts for the reduction of the mode amplitude at the junction position $l=\SI{48}{\um}$ along the resonator, where $k$ is the mode wavevector. Finally, the capacitive coupling term can be expressed in units of the charging energy through the dimensionless parameter $\epsilon = 2C_c \omega Z_c/\pi$, where $C_c$ is the capacity of the JR junction; we estimate $\epsilon \approx 12.5\%$.
\begin{figure}[htbp]
	\begin{center}
		\includegraphics{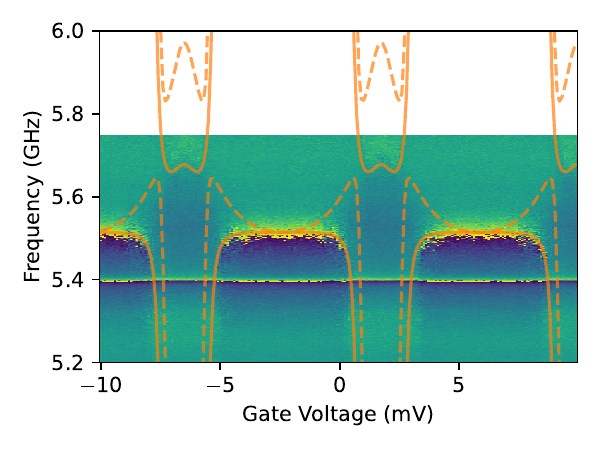}
	\end{center}
	\vspace{-0.4cm}
	\caption{Reflection spectrum of the readout resonator as a function of frequency and gate voltage. The color scale encodes the phase of the reflection coefficient. The lines show the eigenenergies of the Hamiltonian in Eq.~(\ref{eq.readout_H}), calculated for different parameter values (see text). The phase jump at \SI{5.4}{\giga\hertz} is a measurement artifact.
		\label{SI_Fig_Readout_spectrum}}
\end{figure}

In Fig.~\ref{SI_Fig_Readout_spectrum}, we compare the measured readout spectrum with the eigenenergies of $H$. The dashed lines correspond to eigenenergies calculated using the \emph{ab initio} parameter values estimated as described above. The solid lines, which show improved agreement with the experimental data, are obtained using reduced Josephson energies ($\SI{1.45}{\giga\hertz}$), together with smaller effective parameters $\lambda = 0.3$ and $\epsilon = 9.5\%$. In both cases, the frequency $\omega$ of the uncoupled resonator mode is adjusted to reproduce the observed resonance near \SI{5.5}{\giga\hertz}.

\section{Estimation of the photo-assisted tunneling rate and photon to electron conversion efficiency}
The relevant process for single-photon detection, as described in the main text, is the tunneling of an electron across the junction via the absorption of a photon with energy $h\nu$ from the converter mode, while all other electromagnetic degrees of freedom coupled to the junction remain in their ground state. In the limit where the population of the converter mode is small, the photo-assisted tunneling (PAT) process can be treated as an effective single-photon loss channel for this mode. The corresponding loss rate $\kappa_j$ is determined jointly by the junction transparency and its electromagnetic environment. In our device, the electromagnetic environment is defined on one side by the modes of the grAl wire resonator derived in \ref{MicrowaveDesign}, and on the other side by the quantum states of the island. The relevant matrix element for inelastic tunneling corresponds to the process in which a single photon is absorbed from the converter mode while all other modes remain in their ground state. The rate $\kappa_j$ can be therefore be written as~\cite{Stanisavljevic2024}
\begin{equation}
	\begin{split}
		\kappa_j & = |\langle 0|e^{i \lambda_0 (a_0 + a_0^\dagger)}|1\rangle\langle 0|e^{i \lambda_1 (a_1 + a_1^\dagger)}|0\rangle \ldots  \langle 0|e^{i \lambda_m (a_m + a_m^\dagger)}|0\rangle   \langle 0|e^{i \phi/2}|0\rangle  |^2  \ I(V+h \nu  /e)/e \\
		& = \lambda_0^2 e^{-\lambda_0^2} e^{-\lambda_1^2}  \ldots e^{-\lambda_m^2}  |\langle e^{i \phi/2}\rangle  |^2\ I(V+h \nu /e)/e 
	\end{split}
\end{equation}
Here, $a_0$ denotes the annihilation operator of the converter mode, while $a_{m\ge 1}$ correspond to the remaining spectator modes. The displacement amplitudes $\lambda_m$ are set by the vacuum flux fluctuations of each mode multiplied by the elementary charge $e$, and can be expressed in terms of the characteristic impedance of the mode as $\lambda_i^2=\pi Z_c^{(i)}/R_K$ where $R_K=h/e^2$. The last term $|\langle e^{i \phi/2}\rangle|^2$ accounts for the ground state phase fluctuations of the island, which also reduce the tunneling rate. Because the superconducting gap is large ($\Delta \gg \hbar\omega$), PAT processes induce electron tunneling in only one direction, fixed by the applied bias voltage. Consequently, the junction transparency enters the expression for $\kappa_j$ solely through the current–voltage characteristic $I(V)$ of the junction.
\begin{figure}[htbp]
	\begin{center}
		\includegraphics{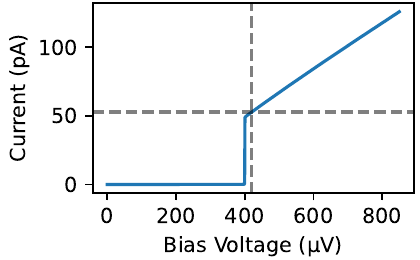}
	\end{center}
	\vspace{-0.4cm}
	\caption{Estimated $I(V)$ characteristics of the converter junction with a normal state resistance of \SI{6.4}{\mega\ohm}. The dashed lines mark the applied voltage bias and the resulting tunnel current used in the expression for the PAT rate $\kappa_j$.}	
	\label{SI_Figure_IV}
\end{figure}

The rate $\kappa_j$ may be rewritten $\kappa_j=\alpha I_T/e$ where $\alpha$ is the product of the different matrix elements and $I_T$ is the tunnel current evaluated at the bias used in the experiment shifted by the energy of the detected photon. We calculate the $I(V)$ characteristic of the converter junction (JC) using the standard BCS density of states for an SIS junction together with its measured large-bias resistance of \SI{6.4}{\mega\ohm}. From this model, we obtain a tunneling current of $I_T = \SI{53}{\pA}$ for a bias of \SI{20}{\uV} above the gap (see Fig.~\ref{SI_Figure_IV}). 

\vspace{5mm}
From the mode table derived in \ref{MicrowaveDesign}, we obtain
\begin{equation}
	\lambda_0^2 e^{-\lambda_0^2} e^{-\lambda_1^2} e^{-\lambda_2^2}\approx 0.26,
\end{equation}
with $\lambda_0 = 0.63$, $\lambda_1 = 0.12$, $\lambda_2 = 0.05$. In order to evaluate $\langle e^{i\phi/2} \rangle$, we neglect the coupling of the island to the readout mode. The Hamiltonian (\ref{eq.readout_H}) then simplifies to the standard Cooper pair box Hamiltonian
\begin{equation}
	H = 4E_C (n - n_0)^2 - E_J \cos\phi,
\end{equation}
The reduction of the quasiparticle tunneling rate is given by the phase-fluctuation factor $|\langle e^{i\phi/2}\rangle|^2$. The factor $1/2$ accounts for the quasiparticle charge being half that of a Cooper pair. Assuming a Gaussian ground state, we use the relation $|\langle e^{i\phi/2} \rangle| = |\langle e^{i\phi}\rangle|^{1/4}$. Close to $n_0 = 0.5$, we obtain $|\langle e^{i\phi}\rangle| = 0.56$ for $E_C/h=E_J/h = 2\pi\times\SI{4.6}{\GHz}$. The final expression of $\kappa_j$ is therefore
\begin{equation}
	\kappa_j =\lambda_0^2 e^{-\lambda_0^2} e^{-\lambda_1^2} e^{-\lambda_2^2} \, |\langle e^{i\phi}\rangle|^{1/2}\, I_T/e
	= 2\pi\times\SI{10.3}{\MHz}.
\end{equation}
Our analysis of the readout spectrum (see \ref{CPB}) shows that the Josephson coupling of the island may be significantly smaller than the \emph{ab initio} values. However, the blockade factor varies slowly with $E_J/E_C$ and remains close to $|\langle e^{i\phi}\rangle| = 0.56$.

\vspace{5mm}
We assume that the dominant loss channels for the converter mode are the junction loss rate $\kappa_j$ and the coupling loss rate $\kappa_c$, neglecting intrinsic losses. In the presence of an incoming photon flux $\Phi$, the steady state mode population is $4 \Phi \kappa_c /(\kappa_c+\kappa_j)^2$. The resulting PAT rate is $4 \Phi \kappa_c \kappa_j /(\kappa_c+\kappa_j)^2 $, which is equal to the incoming photon flux when the matching condition $\kappa_c=\kappa_j$ is verified. Using $\kappa_c=2\pi\times\SI{70}{\mega\hertz}$ (see above), we estimate a photon to electron conversion efficiency of $4 \kappa_c \kappa_j/(\kappa_c+\kappa_j)^2=45\%$.

\section{Measurement setup}
\begin{figure}[htbp]
	\begin{center}
		\includegraphics[width=14 cm]{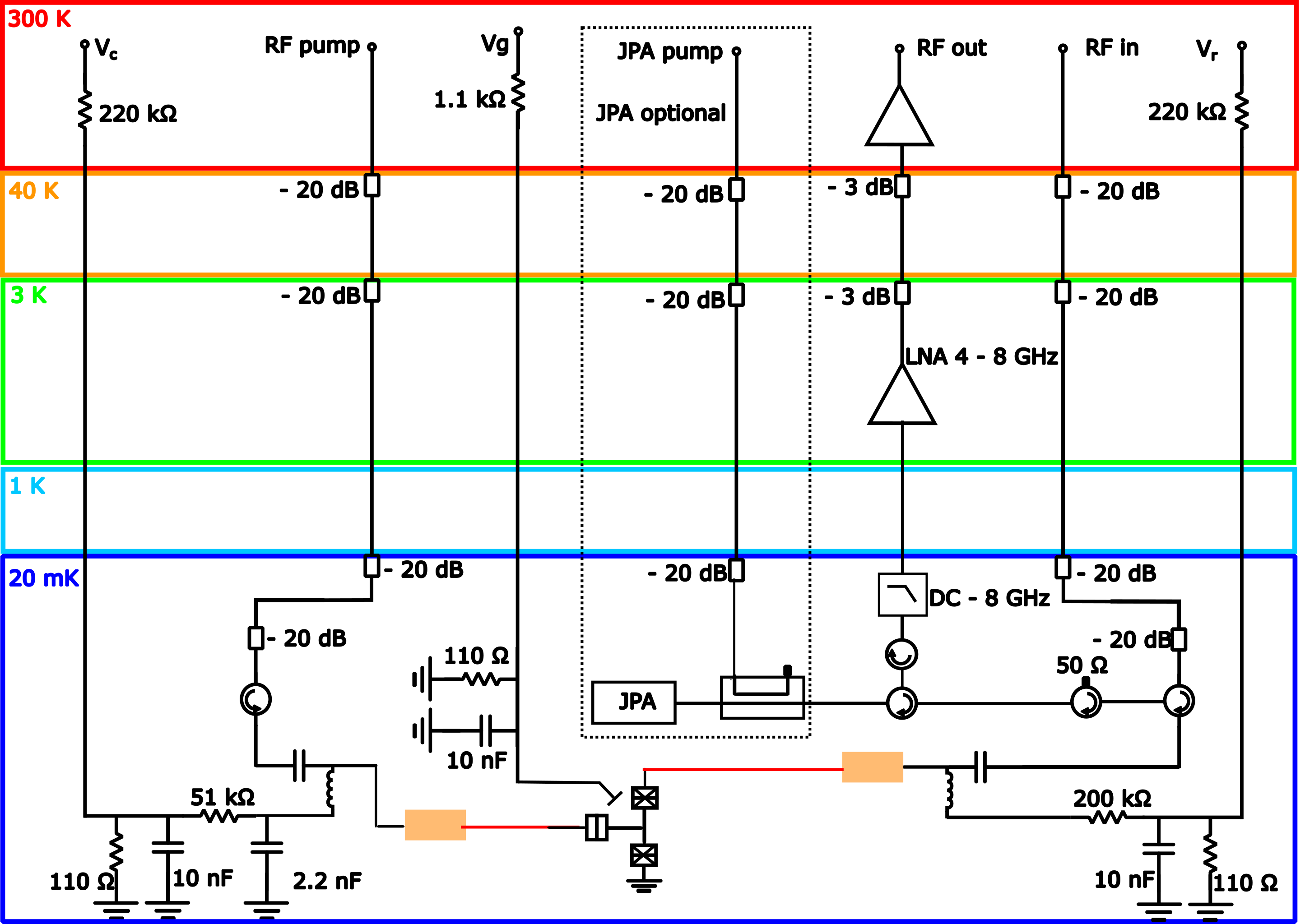}
	\end{center}
	\vspace{-0.4cm}
	\caption{Schematic of the measurement setup cabled in the dilution refrigerator.
	}	
	\label{SI_Figure1}
	\vspace{-0.4cm}
\end{figure}

A schematic of the experimental setup is shown in Fig. \ref{SI_Figure1}.
The device is mounted in a copper sample box equipped with a microwave printed circuit board, indium sealed, and thermally anchored to the mixing-chamber plate of a dilution refrigerator operating at a base temperature of 20 mK. To allow simultaneous dc biasing and microwave measurements, the device is connected to the measurement circuit through a diplexer (or bias tee). Microwave characterization of readout and converter  lines are performed using a vector network analyzer or a modular quantum-control platform. The excitation signal is routed to the sample through an attenuated input line, while the reflected signal is directed to the output line through a series of circulators and amplifiers. The reflected signal is first either amplified by a JPA that works in reflection (the components enclosed by the dashed rectangle in Fig. \ref{SI_Figure1} indicate the optional JPA setup), or directly connected to a low-noise HEMT amplifier at 4 K, followed by two room-temperature amplifiers.
The junctions and the gate are voltage-biased via voltage dividers and filtering capacitors connected to the low-frequency ports of the diplexer/bias tee at base temperature. The biasing circuits involve \SI{50}{\kohm} and \SI{200}{\kohm} series resistor for reliable circuit biasing.

\section{Estimation of the readout error}
\begin{figure}[htbp]
	\begin{center}
		\includegraphics[width=12 cm]{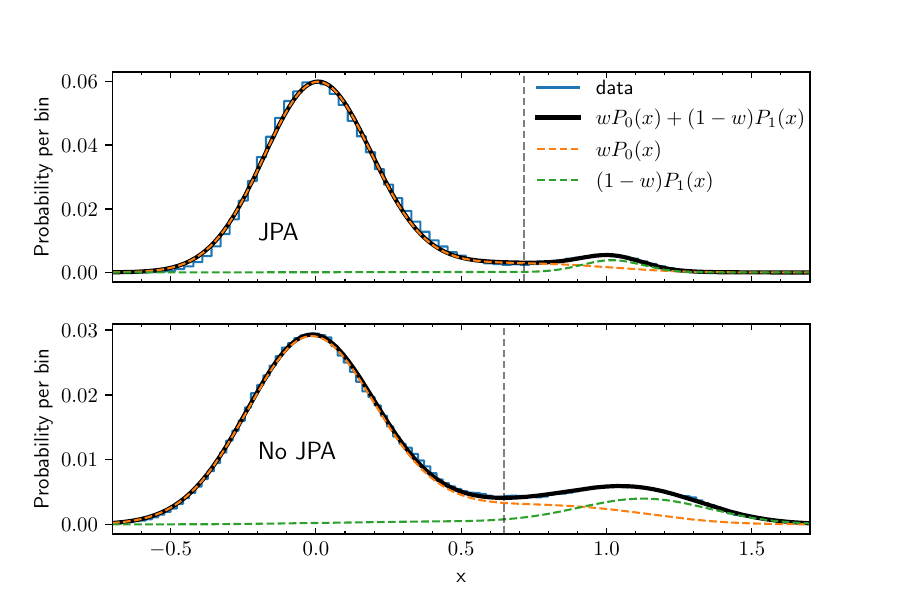}
	\end{center}
	\vspace{-0.4cm}
	\caption{Histograms measured at $V_c=\SI{385}{\uV}$ and $N=0$, obtained with (upper panel) and without (lower panel) the JPA, together with fits using a weighted sum of two sliding Gaussian distributions. The vertical dashed line indicates the threshold set at the local minimum between the two peaks. The measurement time $T$ is \SI{100}{\ns} with the JPA and \SI{200}{\ns} without.
	}	
	\label{SI_Figure2}
	\vspace{-0.4cm}
\end{figure}

As discussed in the main text in connection with the data shown in Figs.~2 and 3, the probability distribution of the measured quadrature at $V_c=\SI{385}{\uV}$, in the absence of any incoming signal at the detector input, is not well described by the sum of two Gaussian functions—one associated with each parity state—as shown in Fig.~\ref{SI_Figure2}. We attribute this to the fact that the parity changes during the measurement time $T$. In order to model this effect, we suppose that the mean signal is $c_0$ ($c_1$) if the parity is even (odd), that the parity switches from $j$ to $i$ at a rate $\gamma_{ij}$ and that the measurement noise is set by the amplifier noise to a rms value $\sigma$. We suppose that only one parity jump can happen during the measurement time. If the initial parity is odd, the probability to obtain the outcome $x$ is a sliding gaussian distribution given by
\begin{align}
	P_1(x) & = \int_0^T \frac{\gamma_{01} \, e^{-\gamma_{01} t}}{\sqrt{2\pi\sigma_1^2}}  e^{-\frac{[x- (t/T)c_1-(1-t/T)c_0 ]^2}{2 \sigma_1^2}}  \, dt + \frac{e^{-\gamma_{01} T}}{\sqrt{2\pi\sigma_1^2}} e^{-\frac{(x-c_1)^2}{2 \sigma_1^2}} \nonumber \\
	&= \frac{\gamma_{01} T}{2(c_0 - c_1)}
	\Bigg\{
	\left[
	\operatorname{erf}\!\left(
	\dfrac{
		c_0^2 - c_0(c_1 + x) + c_1 x - \gamma_{01} \sigma_1^2 T
	}{
		\sqrt{2}\,\sigma_1 (c_0 - c_1)
	}
	\right)
	-
	\operatorname{erf}\!\left(
	\dfrac{
		\gamma_{01} \sigma_1 T 
	}{
		\sqrt{2}\,(c_1 - c_0)
	} + \dfrac{c_1 - x
	}{
		\sqrt{2}\,\sigma_1 
	} 
	\right)
	\right]
	\nonumber\\[1.0ex]
	&\qquad\times
	\exp\!\left[
	\dfrac{
		\gamma_{01} T(\gamma_{01} \sigma_1^2 T - 2(c_0 - c_1)(c_0 - x))
	}{
		2(c_0 - c_1)^2
	}
	\right]
	\Bigg\} +\frac{e^{-\gamma_{01} T}}{\sqrt{2\pi\sigma_1^2}} e^{-\frac{(x-c_1)^2}{2 \sigma_1^2}} .
	\label{eq:slidingclosed}
\end{align}
If the initial parity is even, we obtain the distribution $P_0(x)$ in the same way with $\gamma_{01} \rightarrow \gamma_{10}$, $\sigma_{0} \rightarrow \sigma_{1}$ and $c_0 \leftrightarrow c_1$.
The measured histogram is fitted by a weighted sum $w P_0(x) +(1-w) P_1(x)$, leaving $w$, $c_0$, $c_1$, $\sigma_0$, $\sigma_1$ as free parameters and fixing $\gamma_{01}=\gamma_{10}= 1.06\, \si{\per\micro\second}$ obtained from the pulsed measurement (see Fig.~4 of the main text). In principle, $\sigma_0$ and $\sigma_1$ should be identical, but the data are better fit when they are allowed to differ. Fixing the threshold to $x_{\rm threshold}$, the readout errors are given by
\begin{equation*}
	\epsilon_0 = \int_{x_{\rm threhold}}^{\infty} P_0(x) \, dx  \qquad \
	\epsilon_1 = \int^{x_{\rm threhold}}_{- \infty} P_1(x) \, dx \, .    
\end{equation*}
Figure S2 shows the histogram of the readout signal measured with JPA at $V_c=\SI{385}{\uV}$ in the absence of incoming photons ($N=0$). Setting the threshold at the local minimum between the two peaks, we obtain $w=0.96$, $\epsilon_0=3.2\%$, $\epsilon_1=7.3\%$ with JPA and $w=0.88$, $\epsilon_0=8.6\%$, $\epsilon_1=12.6\%$ without JPA. 

\section{Power calibration}

The calibration of the number of photons per pulse first relies on a quantitative measurement of the output pulse from the microwave source characterized at room temperature using a calibrated spectrum analyzer in zero span mode. Each recorded pulse trace is fitted to a Gaussian profile, from which we extract the pulse energy $E_{\rm pulse}$ as the integral of the fitted function. We then determine the total attenuation $A_{\rm tot}$ of the microwave lines and components through a series of vector network analyzer (VNA) measurements. 
First, the attenuation of the room-temperature section of the setup, including filters, cables, and dc blocks, is measured, yielding the value $A_0$. 
The attenuation of the cryostat cables, $A_1$, is obtained by measuring in series the transmission through the pump line of interest and a nominally identical return line (same length, attenuators, and cable type); $A_1$ is then taken as half of the total measured attenuation. 
Finally, the attenuation at the mixing-chamber stage, $A_2$, is evaluated from direct VNA measurement of the full cryogenic assembly, including circulators, diplexer, attenuators, and cables leading to the chip at room temperature. %
The overall attenuation of the complete line is taken to be $A_{\rm tot}=A_0 + A_1 + A_2 = 90.6~\pm 0.5~\mathrm{dB}$ at $\nu = 9.97~\mathrm{GHz}$.
The average number of photons per pulse at the device input is then obtained from  
\[
N = \frac{E_{\mathrm{pulse}}\times  10^{-A_{\rm tot}/10}}{h\nu}.
\]

\section{Viterbi algorithm}
The island’s charge parity undergoes random transitions between even and odd states at fixed rates $\gamma_{01}$ and $\gamma_{10}$, forming a two-state Markov process. Labeling the parity at times $t_i$ as $q_i$, the likelihood of obtaining a sequence of observations $x_i$ is
\begin{equation}
	P(X|Q)=\prod_{i=1}^T P(x_i|q_i)
\end{equation}
where $T$ is the number of measurements and $X$ and $Q$ denote the two sequences $\{x_i\}$ and $\{q_i\}$, respectively. The conditional probabilities $P(x|0)\equiv P_0(x)$ and $P(x|1)\equiv P_1(x)$ are supposed to be known. Here we approximate them as Gaussian distributions whose parameters are extracted by fitting the histogram of $X$.
The Viterbi algorithm determines the most likely parity sequence $Q$ that generated the observations $X$ \cite{Viterbi1967}. Denoting the binning time by $dt=t_{i+1}-t_i$, the pseudocode of the algorithm is
\vspace{2mm}
\begin{algorithmic}
	\State $w_{00} \gets \exp(-\gamma_{10}*dt)$
	\State $w_{11} \gets \exp(-\gamma_{01}*dt)$ 
	\State $w_{10} \gets 1-w_{00}$ 
	\State $w_{01} \gets 1-w_{11}$
	\State $p_0[1] \gets \gamma_{01}/(\gamma_{01}+\gamma_{10})*P_1(x_1)$
	\State $p_1[1] \gets \gamma_{10}/(\gamma_{01}+\gamma_{10})*P_0(x_1)$
	\State $\operatorname{normalize}(p_0[1], p_1[1])$
	\For{$i=2,3,\ldots, T$}
	\State $p_{00} \gets w_{00}*p_0[i-1] $
	\State $p_{01} \gets w_{01}*p_1[i-1] $
	\If{$p_{01}>p_{00}$}
	\State $p_0[i] \gets p_{01}*P_0(x_i)$
	\State $b[i,0] \gets 1$
	\Else
	\State $p_0[i] \gets p_{00}*P_0(x_i)$
	\State $b[i,0] \gets 0$
	\EndIf
	\State $p_{10} \gets w_{10}*p_0[i-1] $
	\State $p_{11} \gets w_{11}*p_1[i-1] $
	\If{$p_{11}>p_{10}$}
	\State $p_1[i] \gets p_{11}*P_1(x_i)$
	\State $b[i,1] \gets 1$
	\Else
	\State $p_1[i] \gets p_{10}*P_1(x_i)$
	\State $b[i,1] \gets 0$
	\EndIf
	\State $\operatorname{normalize}(p_0[i], p_1[i])$
	\EndFor
	\If{$p_1[T]>p_0[T]$}
	\State $q[T] \gets 1$
	\Else
	\State $q[T] \gets 0$
	\EndIf
	\For{$i=T-1,T-2,\ldots , 1$}
	\State $q[i] \gets b[i+1,q[i+1]]$
	\EndFor
	\Return q
\end{algorithmic}
The purpose of the normalize function is to rescale its arguments so that their sum is equal to unity.

\section{Temperature dependent poisoning}
\begin{figure}[htbp]
	\begin{center}
		\includegraphics[width=10 cm]{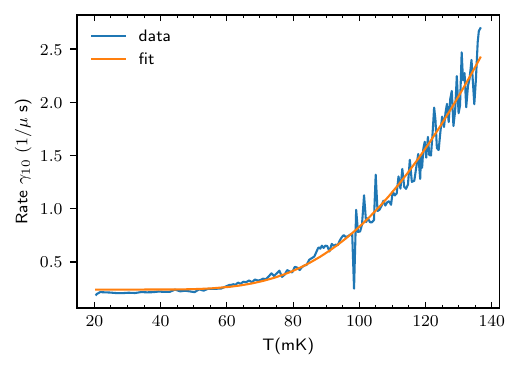}
	\end{center}
	\vspace{-0.4cm}
	\caption{Temperature dependence of the poisoning rate $\gamma_{10}$. The solid orange line is a fit to the data using Eq.~(\ref{eq:nth}).}	
	\label{SI_FigureTemperature}
	\vspace{-0.4cm}
\end{figure}

We assume that the intrinsic losses of the converter resonator are negligible compared with the coupling loss $\kappa_c$ and the PAT rate $\kappa_j$. Under this assumption, the detection efficiency is $\eta = 4\kappa_j \kappa_c/(\kappa_j+\kappa_c)^2$. In the absence of an incoming signal, the steady-state population of the resonator is purely thermal, with an average photon number
$n_{\rm th}=\kappa_c \, (n_{\rm BE}(T)+n_0)/(\kappa_c+\kappa_j)$,
where $n_{\rm BE}(T) = \bigl(e^{h\nu/kT}-1\bigr)^{-1}$ is the Bose--Einstein distribution at frequency $\nu=\SI{10}{\GHz}$, $n_0$ is a residual out of equilibrium population and the junction is modeled as a zero-temperature bath \cite{Stanisavljevic2024}. The resulting poisoning rate is given by
\begin{equation}
	\gamma_{10}
	= \kappa_j n_{\rm th} = \frac{\eta \, \kappa}{4} \left(n_{\rm BE}(T)+n_0 \right).
	\label{eq:nth}
\end{equation}
where $\kappa = \kappa_c+\kappa_j=2\pi \times \SI{81.6}{\MHz}$ is the measured linewidth. Figure~\ref{SI_FigureTemperature} shows the measured evolution of $\gamma_{10}$ as a function of temperature. Fitting the data with Eq.~(\ref{eq:nth}) yields $\eta = 56\%$ and $n_0=3\times 10^{-3}$.

\section{Non-linear photo-assisted tunneling processes}
\begin{figure}[htbp]
	\begin{center}
		\includegraphics[width=10 cm]{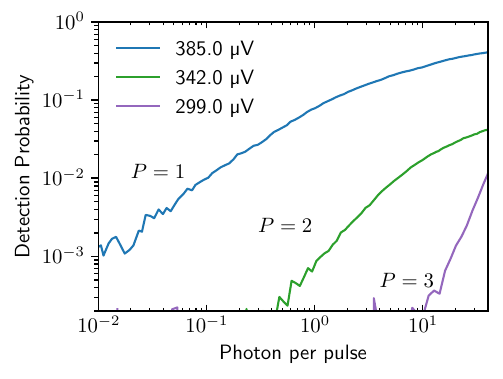}
	\end{center}
	\vspace{-0.4cm}
	\caption{Detection probability as a function of the average microwave photon per pulse taken for 3 different subgap voltages corresponding to linear ($P=1$) and non-linear processes ($P=2,3$).}	
	\label{SI_FigureP123}
	\vspace{-0.4cm}
\end{figure}

In addition to the linear conversion process discussed in the main text at a bias of \(V_c = 385~\mu\mathrm{V}\), higher-order photo-assisted tunneling mechanisms involving the absorption of \(P = 2\) and \(3\) photons become visible at lower bias voltages. 
We present in figure~\ref{SI_FigureP123} corresponding data, acquired without JPA and using similar parameters as figure 2 of the main text, that shows the microwave power dependence of these PAT steps. The different slopes observed for each curve reflect the non-linear nature of the underlying multiphoton processes \cite{Stanisavljevic2024}. 

\section{Readout power optimization with and without JPA}
\label{Backaction}
\begin{figure}[htbp]
	\begin{center}
		\includegraphics[width=10 cm]{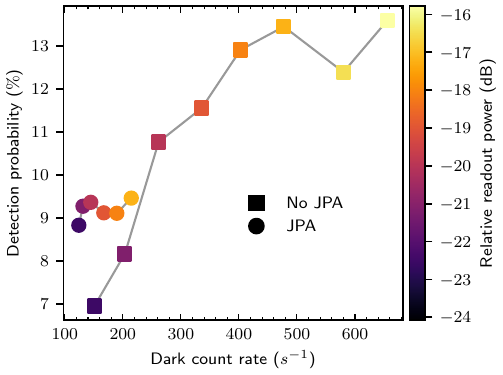}
	\end{center}
	\vspace{-0.4cm}
	\caption{Detection probability of a charge parity jump induced by a microwave pulse containing, on average, one photon, as a function of the dark count rate, measured with and without a JPA. The data points correspond to different powers of the continuous readout tone, indicated by the color scale. The data analysis follows the procedure described in Fig.~4 of the main text.}	
	\label{SI_FigureJPANoJPA}
	\vspace{-0.4cm}
\end{figure}

Optimization of the detector performance relies on tuning the microwave readout power to achieve the desired trade-off between detection probability and dark count rate. Figure~\ref{SI_FigureJPANoJPA} shows the detection probability for pulses containing, on average, one photon, plotted as a function of the dark count rate for measurements performed with and without a Josephson parametric amplifier (JPA). Each data point corresponds to a different readout power, increasing from left to right according to the color scale.

Without the JPA, increasing the readout power substantially improves the detection probability by reducing the detection error; however, it simultaneously leads to a strong increase in the dark count rate, as the readout tone increasingly photo-assists tunneling events onto the island. In contrast, when the JPA is employed, the detector operates in a more favorable regime. Detection errors no longer significantly limit the detection probability, enabling high detection probabilities to be achieved at low readout power and therefore at low dark count rates.

\end{document}